%% file: sourcemod_main_arxiv.tex
\documentclass[10pt, twocolumn, superscriptaddress, floatfix, showkeys, amsmath, amssymb, prb]{revtex4-1}

\usepackage{enumitem}
\setlist[enumerate]{itemsep=0mm}
\usepackage{comment}
\usepackage[normalem]{ulem}
\usepackage{setspace}
\usepackage{framed,graphicx,xcolor}
\usepackage{float}
\setlist{nolistsep}
\usepackage{multirow}

\begin{document}

\title{Sparsity Enables Estimation of both Subcortical and Cortical Activity from MEG and EEG}

\author{Pavitra Krishnaswamy}
\affiliation{Athinoula A. Martinos Center for Biomedical Imaging, Department of Radiology, Massachusetts General Hospital, Charlestown, MA 02129, USA}
\affiliation{Harvard-MIT Division of Health Sciences and Technology, Cambridge, MA 02139, USA}
\affiliation{Institute for Infocomm Research, A{*}STAR, Singapore 138632, Singapore}
\author{Gabriel Obregon-Henao}\affiliation{Athinoula A. Martinos Center for Biomedical Imaging, Department of Radiology, Massachusetts General Hospital, Charlestown, MA 02129, USA}
\author{Jyrki Ahveninen}\affiliation{Athinoula A. Martinos Center for Biomedical Imaging, Department of Radiology, Massachusetts General Hospital, Charlestown, MA 02129, USA}
\author{Sheraz Khan}
\affiliation{Athinoula A. Martinos Center for Biomedical Imaging, Department of Radiology, Massachusetts General Hospital, Charlestown, MA 02129, USA}
\affiliation{Department of Neurology, Massachusetts General Hospital, Charlestown, MA 02129, USA}
\affiliation{Harvard Medical School, Boston, MA 02115, USA}
\author{Behtash Babadi}
\affiliation{Athinoula A. Martinos Center for Biomedical Imaging, Department of Radiology, Massachusetts General Hospital, Charlestown, MA 02129, USA}
\affiliation{Department of Electrical \& Computer Engineering, University of Maryland, College Park, MD 20742, USA}
\author{Juan Eugenio Iglesias}\affiliation{Athinoula A. Martinos Center for Biomedical Imaging, Department of Radiology, Massachusetts General Hospital, Charlestown, MA 02129, USA}
\author{Matti S. Hamalainen}
\email{msh@nmr.mgh.harvard.edu}
\affiliation{Athinoula A. Martinos Center for Biomedical Imaging, Department of Radiology, Massachusetts General Hospital, Charlestown, MA 02129, USA}
\affiliation{Department of Neuroscience and Biomedical Engineering, Aalto University School of Science, Espoo 02150, Finland}
\affiliation{NatMEG, Department of Clinical Neuroscience, Karolinska Institute, Stockholm 17177, Sweden}
\author{Patrick L. Purdon}
\email{patrickp@nmr.mgh.harvard.edu}
\affiliation{Department of Anesthesia, Critical Care and Pain Medicine, Massachusetts General Hospital, Boston, MA 02114, USA}
\affiliation{Harvard Medical School, Boston, MA 02115, USA}

\date{\today}

\begin{abstract}
Abstract: Subcortical structures play a critical role in brain function. However, options for assessing electrophysiological activity in these structures are limited. Electromagnetic fields generated by neuronal activity in subcortical structures can be recorded non-invasively using magnetoencephalography (MEG) and electroencephalography (EEG). However, these subcortical signals are much weaker than those due to cortical activity. In addition, we show here that it is difficult to resolve subcortical sources, because distributed cortical activity can explain the MEG and EEG patterns due to deep sources. We then demonstrate that if the cortical activity can be assumed to be spatially sparse, both cortical and subcortical sources can be resolved with M/EEG. Building on this insight, we develop a novel hierarchical sparse inverse solution for M/EEG. We assess the performance of this algorithm on realistic simulations and auditory evoked response data and show that thalamic and brainstem sources can be correctly estimated in the presence of cortical activity. Our analysis and method suggest new opportunities and offer practical tools for characterizing electrophysiological activity in the subcortical structures of the human brain.
\end{abstract}


\maketitle

\definecolor{shadecolor}{gray}{0.95}

\input{srcmod_intro.tex}
\input{srcmod_theory.tex}
\input{srcmod_algorithm.tex}
\input{srcmod_datares.tex}
\input{srcmod_discussion.tex}
\input{srcmod_methods.tex}

\vspace{1ex}
\textbf{Acknowledgments:} We acknowledge data collection assistance from Samantha Huang, Stephanie Rossi, and Tommi Raij; helpful discussions on source space construction with Koen Van Leemput, Imam Aganj, Doug Greve and Bruce Fischl at the MGH/HST Athinoula A. Martinos Center for Biomedical Imaging; and helpful discussions on canonical correlations and sparsity with Demba Ba and Emery N. Brown at the Massachusetts Institute of Technology. This work was supported by NIH grants P41-EB015896, 5R01-EB009048, and NIH grant 1S10RR031599-01 to M.S.H, and NIH grant DP2-OD006454 to P.L.P.

\vspace{1ex}
{\textbf{Author Contributions:}} P.K.S, M.S.H and P.L.P designed research; P.K.S, G.O.H and S.K performed simulations and calculations; P.K.S, J.A and S.K collected experimental datasets; P.K.S, G.O.H, P.L.P and M.S.H contributed new analytic methods with inputs and tools from B.B and E.I; P.K.S and G.O.H analyzed the data; P.K.S, M.S.H and P.L.P wrote the manuscript. All authors approved the final manuscript.

\input{sourcemod_main_arxiv.bbl}
\end{document}

%% file: srcmod_intro.tex
\vspace{-3ex}

\section*{Introduction}

Deep brain structures play important roles in brain function. For example, brainstem and thalamic relay nuclei have a central role in sensory processing \cite{Jones2001,Jones2002}. Thalamocortical and hippocampal oscillations govern states of sleep, arousal, and anesthesia \cite{Steriade1993a, Buzsaki2016, Ching2010, Steriade1996, Crunelli2010, Hughes2005}. Striatal regions are crucial for movement planning, while limbic structures like the hippocampus and amygdala drive memory, emotion and learning \cite{Graybiel2000, Alexander1986, Haber2003, Amygdala1, Learning1}. Further, altered signaling within the thalamus, striatum, hippocampus, and amygdala underlies pathologies such as autism, dementia, and depression \cite{Blumenfeld2010}. Much of our understanding of the function of subcortical structures comes from lesion studies and invasive electrophysiological recordings in animal models. Improved tools to characterize subcortical activity in humans would make it possible to analyze interactions between subcortical structures and other brain areas, and could be used to better understand how subcortical activity relates to perception, cognition, behavior and associated disorders.

At present, techniques for characterizing deep brain dynamics are limited. Invasive electrophysiological recordings \cite{NeidermeyerErnstDaSilva} in humans are generally limited to regions that need to be monitored for clinical purposes. Functional magnetic resonance imaging (fMRI) can non-invasively assess activity deep in the brain with a spatial resolution of up to $\sim1-2\,{\rm mm}$, but cannot record fast signals or oscillations. Magnetoencephalography (MEG) and electroencephalography (EEG) non-invasively measure fields generated by neural currents with millisecond-scale temporal resolution \cite{Hamalainen1993}, and M/EEG source estimation \cite{Baillet2001} is widely used to spatially resolve neural dynamics to within $\sim0.5-2\,{\rm cm}$ in the cerebral cortex \cite{Dale1993a, Dale2000, Ou2009, Gramfort2013}. However, it remains an open question as to whether M/EEG can be used to estimate neural currents in deep brain structures.

\begin{figure}[!hb]
\noindent \begin{centering}
\includegraphics[width=3.42in]{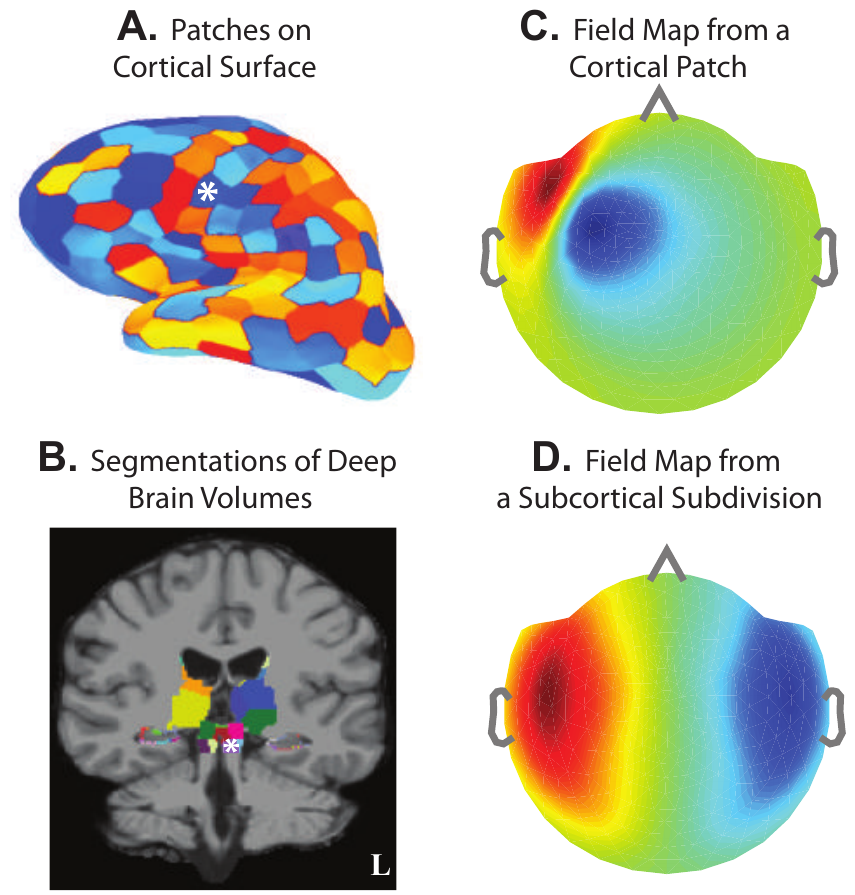}
\par\end{centering}

\caption[Illustration of Source Spaces and Field Patterns]{ \textbf{Illustration of Source Spaces and Field Patterns.}(\emph{A}) Cortical source space $\mathcal{C}$ comprising patches (sized ~$650\,{\rm mm}^{2}$) on the gray-white matter surface interface. \emph{(B)} Subcortical source space $\mathcal{S}$ comprising volume subdivisions (sized $175-1800\,{\rm mm^{3}}$) in the caudate, putamen, amygdala, thalamus, brainstem, and surface patches (sized $47\,{\rm mm^{2}}$) on hippocampus.\emph{ }(\emph{A-B}) The set of cortical and subcortical divisions $\mathcal{B}=\mathcal{C}\,\cup\,\mathcal{S}$ form the full distributed source space.\emph{ (C-D)} Example noiseless MEG field pattern arising from activity in a frontotemporal cortical patch\emph{ }and a brainstem subdivision, respectively (white stars). Inflated surfaces and field maps have a left-right convention opposite to the MRI views.\emph{ (C vs. D)} Cortical fields tend to have more focal spatial patterns, while subcortical fields tend to be more distributed.
\label{fig:srcspace}}
\end{figure}

\begin{figure}[!hb]
\noindent \begin{centering}
\includegraphics[width=3.42in]{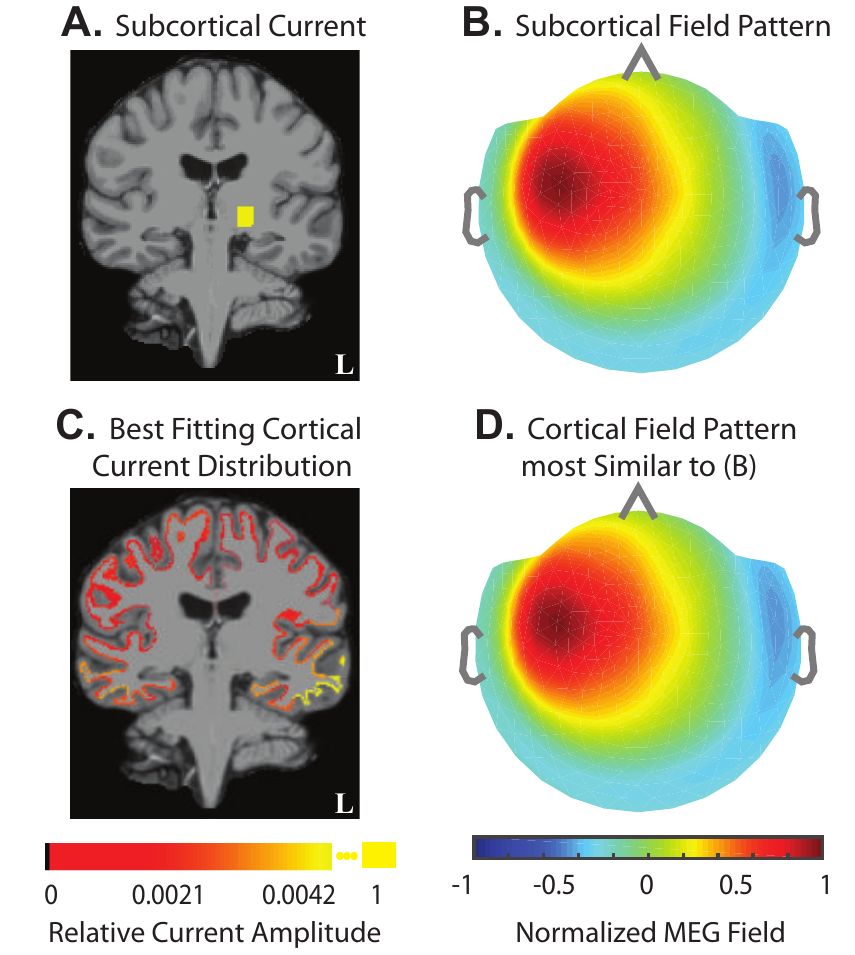}
\par\end{centering}

\caption[Algorithm: Challenge Illustration]{\textbf{ Fields Generated by Subcortical Sources can be Explained
by Currents on the Cortex}. (\emph{A}-\emph{B}) Example unit source current in left ventroposterolateral (VPL) thalamus (sized $1.5\,{\rm cm^{3}}$), and the corresponding noiseless MEG field pattern. \emph{(}C-\emph{D)} Distribution of currents on cortical surface patches (sized $650\,{\rm mm^{2}}$) that reproduce the MEG field pattern generated by the subcortical source. Panels \emph{C} and \emph{D} show the fitted cortical currents and MEG field pattern. The source current plots in \emph{ (A vs. C)}  are the resultant currents from dipoles within a subdivision.  The field maps in \emph{ (B and D)} are normalized. The fitted cortical field pattern is indistinguishable from the simulated subcortical pattern. This analysis illustrates how a subcortically generated field can be explained by some distribution of cortical currents.
\label{fig:challenge}}
\end{figure}

The anatomy of deep brain structures poses two significant challenges for source estimation with M/EEG. First, deep brain structures are farther away from the sensors than the cerebral cortex and thus produce lower-amplitude M/EEG signals than the cortex. A second, perhaps more fundamental problem stems from the fact that subcortical structures are enclosed by the cortical mantle. Thus, measurements arising from activity within deep brain structures can potentially be explained by a surrogate distribution of currents on the cortical surface. This ambiguity would also mean that it is harder to estimate subcortical activity when cortical activity is occurring simultaneously.

We reason, however, that the above challenges could be mitigated if only a limited number of cortical sites have activity together with subcortical structures. In many neuroscience studies, salient cortical activity at any moment in time tends to be restricted to a small set of well-circumscribed areas. It follows that if we could identify this sparse subset of active cortical sources (e.g., as in \cite{Babadi2014}) and prune away the remaining irrelevant cortical sources, we might have a chance at recovering the locations and time courses of both cortical and subcortical sources. The feasibility of this approach would depend upon the degree of overlap amongst the M/EEG field patterns of these candidate sources and the signal-to-noise ratio of the measurements.

In this paper, we analyze the M/EEG field patterns due to cortical and subcortical sources and assess the extent to which sparse cortical and subcortical sources can be distinguished. We then introduce a hierarchical sparse estimation algorithm to characterize both cortical and subcortical activity. Finally, we demonstrate the algorithm’s performance on simulated and experimental M/EEG data containing both cortical and subcortical activity. 

%% file: srcmod_theory.tex
\vspace{-4ex} 

\section*{Theory}

\textbf{Neural Sources and M/EEG Fields}: Primary neural currents \cite{HariIlmoniemi1986}, usually modeled with an ensemble of current dipoles, generate M/EEG fields $\mathbf{Y}$:
\begin{equation}
\mathbf{Y}_{{\scriptstyle {\scriptscriptstyle N\times T}}}=\mathbf{G}_{{\scriptscriptstyle N\times M}}\mathbf{X}_{{\scriptscriptstyle M\times T}}+\mathbf{V}_{{\scriptscriptstyle N\times T}},\label{eq:meas_model}
\end{equation}
where $\mathbf{G}$ is the gain matrix determined by the quasistatic approximation of Maxwell's equations, $\mathbf{X}$ contains the amplitudes of the current dipole sources, $\mathbf{V}$ is the  noise, $N$ is the number of sensors, $M$ is the number of sources, and $T$ is the number of time points measured. To simplify notation, we assume that the data, the gain matrix, and the noise in Eq.\ref{eq:meas_model} have been whitened to account for the spatial covariance $\mathbf{Q}_{{\scriptscriptstyle N\times N}}$ of the actual observation noise \cite{Engemann2015}, so that $\mathbf{V}$ is Gaussian with zero mean and identity covariance matrix $\mathbf{I}_{N\times N}$. When $T=1$, we use notation $\mathbf{y}$ and $\mathbf{x}$ in lieu of $\mathbf{Y}$ and $\mathbf{X}$.

We employ high-resolution structural MRIs from healthy volunteers to delineate cortical surfaces and subcortical anatomic regions  to define the locations and orientations of the elementary dipole sources (Methods). We place sources on cortical surfaces and in subcortical volumes, and cluster proximal groups of dipoles into surface patches or volume subdivisions sized to homogenize signal strengths (Methods, Fig.\ref{fig:srcspace}\emph{A}-\emph{B}, SI. I). The resulting set $\mathcal{B}$ of $K$ patches and subdivisions, together called divisions, constitutes the distributed brain source space.

We can then group the columns of $\mathbf{G}$ and rows of $\mathbf{X}$ according to these divisions, and rewrite Eq.\ref{eq:meas_model} as:
\begin{equation}
\mathbf{Y}_{{\scriptscriptstyle {\scriptscriptstyle N\times}T}}=\sum_{k=1}^{K}\mathbf{G}_{k}\mathbf{X}_{k}+\mathbf{V}_{{\scriptscriptstyle N\times T}},
\end{equation}
where $\mathbf{G}_{k}$ and $\mathbf{X}_{k}$ denote the gain matrix and source currents within the $k^{{\rm th}}$ division respectively. We compute $\mathbf{G}_{k}$ for each division $k$ in $\mathcal{B}$ (Methods). We denote gain matrices and source currents for a set of divisions $\mathcal{F\subset\mathcal{B}}$ by $\mathbf{G}_{\mathcal{F}}=\{\mathbf{G}_{k}\}\,\mathbf{{\rm and}\,X}_{\mathcal{F}}=\{\mathbf{X}_{k}\} \, \forall k \in \mathcal{F}$ respectively. Fig.\ref{fig:srcspace}\emph{C}-\emph{D} illustrate field patterns for one cortical and one subcortical division.

\textbf{Fields Generated by Subcortical Sources can be Explained by Currents on the Cortex}: To analyze distinctions between subcortical and cortical fields, we investigated the extent to which MEG field patterns arising from subcortical currents can be explained by cortical surrogates.

We simulated field pattern $\mathbf{y}_{\mathcal{V}}$ corresponding to unit current in the ventral posterior lateral (VPL) thalamus (Fig.\ref{fig:challenge}\emph{A}-\emph{B}), and assessed if $\mathbf{y}_{\mathcal{V}}$ could be explained by some distribution of cortical source currents. Specifically, we fitted the subcortical field pattern with cortical sources, i.e., we computed the cortical minimum-norm estimate to explain $\mathbf{y}_{\mathcal{V}}$. We found that the resulting currents are small and broadly distributed across several cortical patches (Fig.\ref{fig:challenge}\emph{C}). Further, the goodness-of-fit between the field pattern for the cortical estimate (Fig.\ref{fig:challenge}\emph{D}) and $\mathbf{y}_{\mathcal{V}}$ was $98.5\%$ showing that $\mathbf{G}_{\mathcal{V}}$ can be explained by some combination of sources in the full dense cortical space.

\begin{figure*}[!ht]
\noindent \begin{centering}
\includegraphics[width=7in]{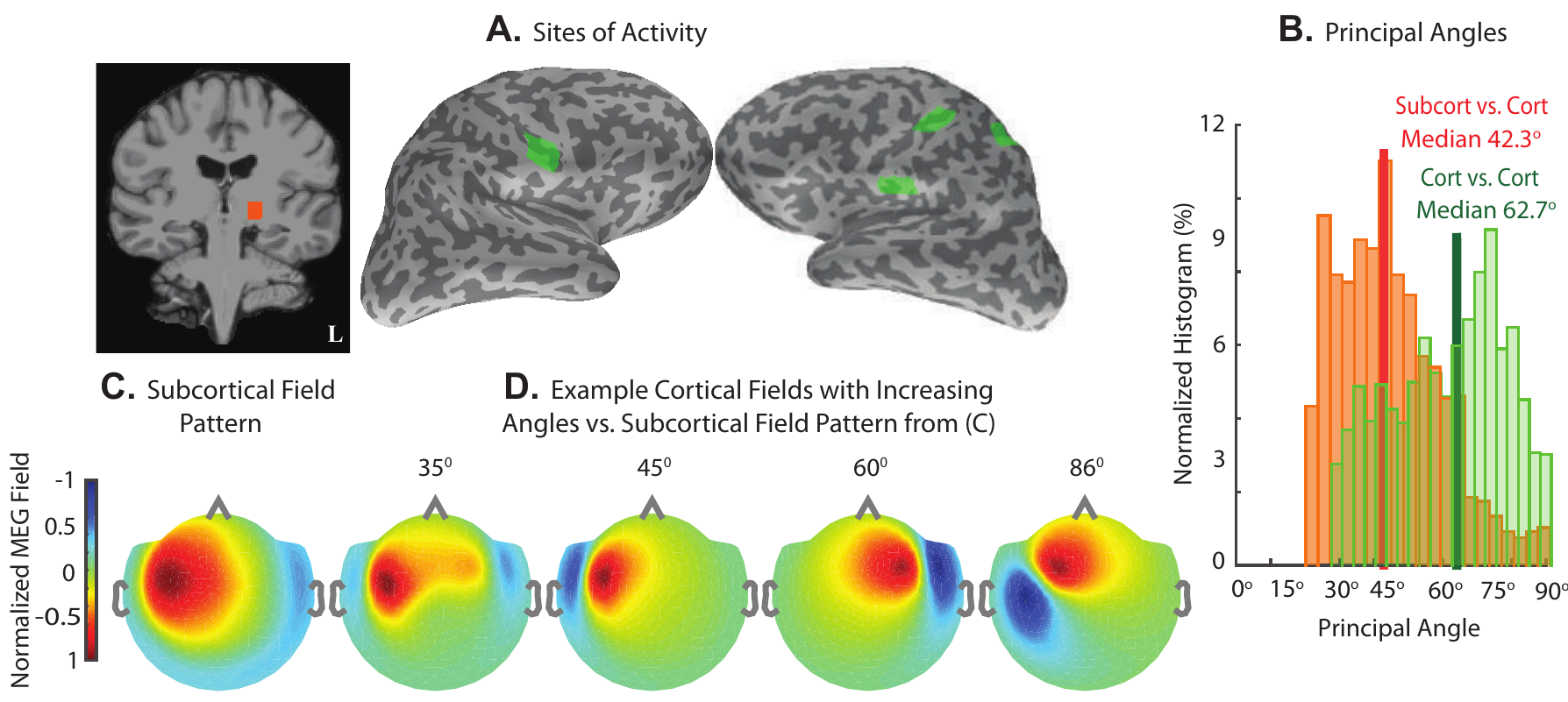}
\par\end{centering}

\caption[Sparsity: Forward Solutions]{\textbf{ Sparsity Makes it Possible to Distinguish Fields from Subcortical
and Cortical Sources}. (\emph{A}) Sources of activity during stimulation of the right median-nerve: left VPL thalamus, primary and secondary somatosensory areas (S1, bilateral S2) and posterior parietal cortex (PPC).\emph{ (B)} Normalized histogram of principal angles, which quantify the correlation between fields arising from all combinations of activity within this neurophysiological source space. The orange histogram shows the distribution of $2992$ subcortical vs. cortical angles, while the green histogram shows the distribution of $11496$ cortico-cortical angles. \emph{(C)} MEG field pattern resulting from activity in VPL thalamus. \emph{(D)} Fields from example cortical source sets, whose gain matrices have indicated principal angles with the subcortical gain matrix, that best fit the subcortical field pattern in \emph{C}. All field map colorscales are normalized to emphasize spatial patterns. The spatial profiles of the cortical and subcortical MEG field patterns are distinct, even for principal angle $30^{\circ}$, and substantially so for principal angles $>45^{\circ}$. These distinctions suggest the feasibility of resolving simultaneous subcortical and cortical activity. \label{fig:sparsity_fwds}}
\end{figure*}

\textbf{Analysis of Forward Solutions}: To generalize the above result, we used principal angles \cite{Bjorck1973,Knyazev2008} to characterize the relationship between the cortical and subcortical field patterns. Principal angles quantify the correlation between linear subspaces, in this case the space spanned by MEG fields arising from sources in different cortical and subcortical regions. For the example in Fig.\ref{fig:challenge}, the maximum principal angle between subspaces spanned by the VPL and cortical gain matrices was $8.2\times10^{-14\,\,\circ}$. The maximum principal angle between any subcortical gain matrix and the cortical gain matrix was $1.5\times10^{-13\,\,\circ}$. We conclude that the presence of the full cortical source space makes it impossible to unambiguously estimate currents in deeper subcortical sources.

\textbf{Sparsity Makes it Possible to Distinguish Fields from Subcortical and Cortical Sources}: We next studied the extent of subspace correlation when subcortical sources are active together with only a small subset of cortex. As an example, we examined a neurophysiologically plausible scenario of median-nerve somatosensory stimulation, which elicits activity in VPL thalamus, primary and secondary sensory cortices (S1, S2), and posterior parietal cortex (PPC) \cite{Forss1994}, and analyzed forward solutions for a source space encompassing these regions (Fig.\ref{fig:sparsity_fwds}\emph{A}).

Specifically, we considered all possible configurations of subcortical and cortical activity in these divisions (Methods), computed principal angles between the subcortical and cortical gain matrices corresponding to each possible configuration, and plotted the distribution of angles (Fig.\ref{fig:sparsity_fwds}\emph{B}). A large proportion of the principal angles are high (median $42.3^{\circ}$), indicating that many different configurations of activity within the sparse cortical and subcortical divisions can be distinguished from one another. We also computed principal angles for all mutually exclusive configurations of activity within the cortical divisions in Fig.\ref{fig:sparsity_fwds}\emph{A}, and found comparable principal angles (median $62.7^{\circ}$). Therefore, in principle, the problem of distinguishing subcortical sources from sparse cortical sources is similar in difficulty to that of distinguishing sparse cortical sources from one another. We also illustrate typical subcortical and cortical field patterns (Methods) for source current configurations corresponding to the various angles in this distribution (Fig.\ref{fig:sparsity_fwds}\emph{C}-\emph{D}). Subcortical and cortical field patterns with principal angles as low as $45{}^{\circ}$ are clearly distinguishable.

This example represents a conservative scenario, but illustrates an approach to characterize the extent to which subcortical sources can be resolved for any given cortical source distribution. We also found similar trends for other more general cases (SI. II). Together, these analyses lead us to conclude that spatial sparsity constraints can enable distinctions between cortical and subcortical field patterns. Based on this analysis, we introduce and test an inverse algorithm that employs a sparse cortical representation to achieve localization of simultaneous subcortical and cortical activity. 

%% file: srcmod_algorithm.tex
\vspace{-2ex}

\section*{Inverse Algorithm}
\textbf{Electromagnetic Inverse Problem}: The electromagnetic inverse problem is to estimate source currents $\mathbf{X}$ underlying M/EEG measurements $\mathbf{Y}$, given the forward gain matrix $\mathbf{G}$ for divisions distributed across the brain. This inverse problem is commonly solved using the linear $\mathit{l}_{2}$ minimum-norm estimator (MNE):
\vspace{-1ex}
\begin{equation}
\mathbf{\hat{X}}(\mathbf{G},\,\mathbf{Y})=\mathbf{W}^{{\rm MNE}}(\mathbf{G})\mathbf{Y}\;,\label{eq:invsoln}
\end{equation}
where $\mathbf{\hat{X}}(\mathbf{G},\,\mathbf{Y})$ is an estimate of $\mathbf{X}$, and the MNE estimator $\mathbf{W}^{{\rm MNE}}$ is a function of $\mathbf{G}$. The performance of $\mathbf{W}^{{\rm MNE}}$ can be assessed using the resolution matrix:
\begin{equation}
\mathbf{\mathbf{K(G})=W}^{{\rm MNE}}(\mathbf{G})\mathbf{G}\label{eq:resmatrix}
\end{equation}
The ideal $\mathbf{K}=\mathbf{I}$ corresponds to a $\mathbf{W^{{\rm MNE}}}$ which exactly recovers the current locations and amplitudes, in the absence of noise \cite{Liu1998}. In practice, source estimates are biased and more extended than the true sources. This non-ideal behavior can be analyzed using the spatial dispersion (SD) and dipole localization error (DLE) metrics (\cite{Molins2008}, see Methods), which indicate how far the inverse solution for a given source spreads from the actual source location. We analyze the performance of the MNE on the subcortical and cortical source estimation problems, and describe our new inverse algorithm.

\begin{figure*}[!ht]
\noindent \begin{centering}
\includegraphics[width=5.5in]{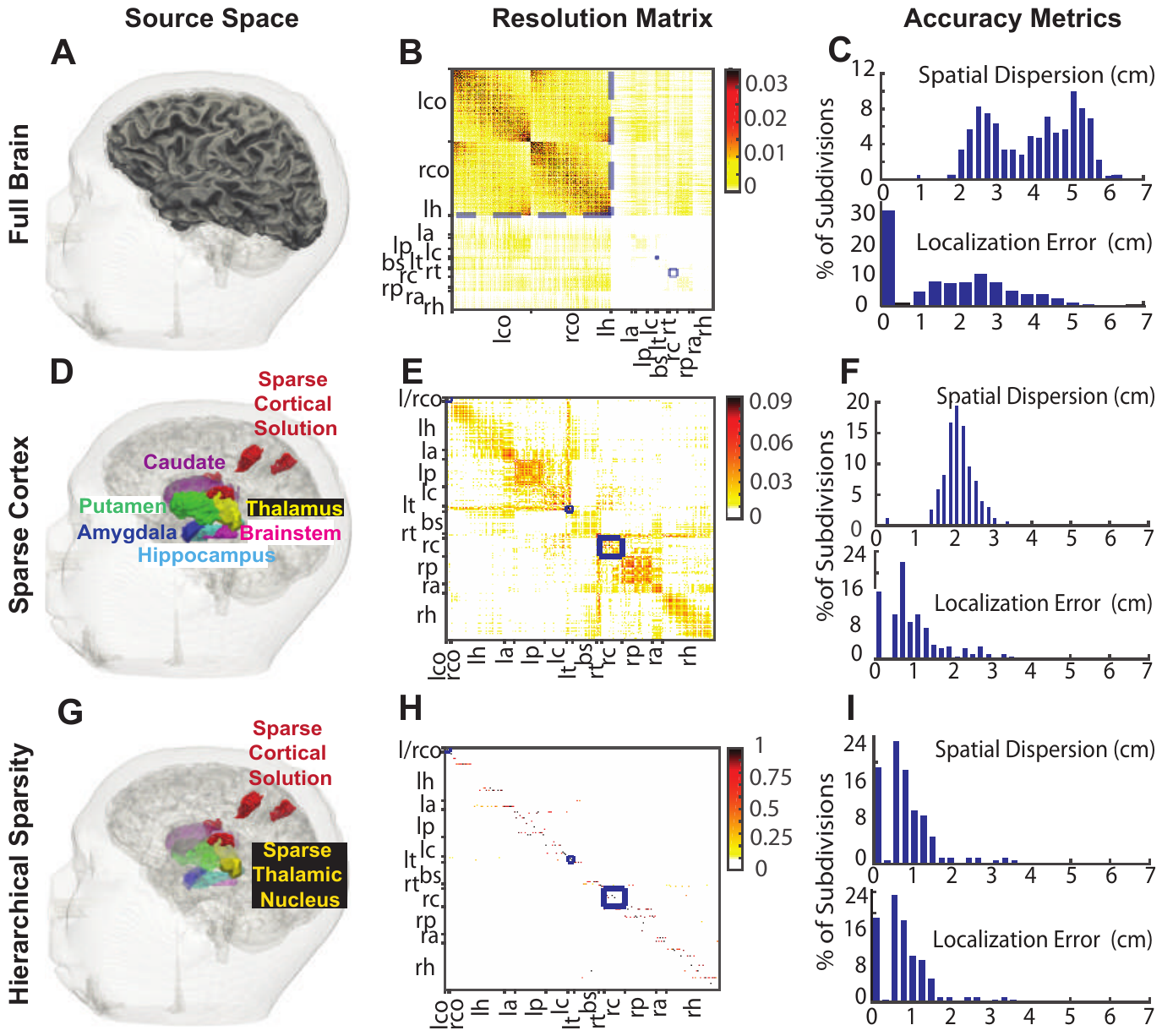}
\par\end{centering}

\caption[Hierarchical Sparsity-Based Algorithm for Subcortical Source Estimation]{\textbf{An Analysis of How Sparsity and Hierarchy Influence Subcortical Source Estimation}. (\emph{A}) Illustration of all brain divisions considered. \emph{(B) }MNE resolution matrix for  the source space in \emph{A}. \emph{(C) }Summary dispersion and error metrics for the resolution matrix  in \emph{B. }Cortical estimates concentrate around the diagonal (low localization error), whereas subcortical estimates spread significantly to the cortex (high spatial dispersion). \emph{(D)} A reduced space comprised of sparse cortical regions that generate somatosensory evoked potentials and all subcortical volumes. \emph{(E-F)} MNE resolution matrix and associated performance metrics for the reduced source space in \emph{D}. The sparse subset of the cortical source space allows subcortical activity to be estimated, albeit with significant spread to non-diagonal regions. \emph{(G)} Final sparse cortical and subcortical source regions identified using an inverse solution employing sparsity constraints. The faded subcortical regions show the hierarchically reduced subcortical source space, while the foreground subcortical regions show estimated sources in the thalamus. \emph{(H-I)} Empirical resolution matrix (one active source per column), and associated performance metrics for sparse solution. Estimates mostly concentrate on and around the diagonal for both cortical and subcortical sources. \emph{(B vs. E vs. H)} Legend: cortex (l/rco), hippocampus (r/lh), amygdala (r/la), putamen (r/lp), caudate (r/lc), thalamus (r/lt), and brainstem (bs). The blue boxes are used to delineate the position of the cortical, left thalamic, and right caudate sources in the resolution matrices. Changes in colorscale range highlight the 3-10x increase in contrast observed when sparse estimation is applied across progressively refined hierarchies. Overall, hierarchical sparsity enables focal spatial resolution with minimal dispersion (or point spread) for inverse solutions incorporating both cortical and subcortical sources.
\label{fig:sparsity-inv}}
\end{figure*}

\textbf{Estimation Performance - Distributed Cortical and Subcortical Sources}: We first considered the problem of estimating neural currents in a set of divisions $\mathcal{B}$ distributed across both cortical and subcortical structures in the brain (Fig.\ref{fig:sparsity-inv}\emph{A}). Fig.\ref{fig:sparsity-inv}\emph{B} shows the MNE resolution matrix $\mathbf{K}^{{\rm MNE}}(\mathbf{G}_{\mathcal{B}})$. We found that estimates for the cortical sources concentrate around the diagonal (Fig.\ref{fig:sparsity-inv}\emph{B}, upper left), implying good resolution for cortical sources. On the other hand, estimates for subcortical sources have low amplitudes on the diagonal, and instead spread to cortical sources (Fig.\ref{fig:sparsity-inv}\emph{B}, upper right and lower left). Fig.\ref{fig:sparsity-inv}\emph{C} shows the distribution of SD and DLE across all divisions. The median SD is $4.23\,{\rm cm}$ (close to the radius of the human brain), and the median DLE is $1.89\,{\rm cm}$. The cortical DLE is approximately $0.5\,{\rm cm}$ and the subcortical DLE is in the $2-3\,{\rm cm}$ range. These findings are consistent with our principal angle analyses (Fig.\ref{fig:challenge}).

\textbf{Estimation Performance - Sparse Cortical and Distributed Subcortical Sources}: Earlier we found the principal angles improve when only sparse subsets of cortical sources are active alongside deep sources (Fig.\ref{fig:sparsity_fwds}). Therefore, we assessed whether the resolution of the MNE inverse solution improves similarly. We considered the somatosensory stimulation example from Fig.\ref{fig:sparsity_fwds}, and constructed a composite source space $\mathcal{B}_{r}$ comprising the sparse cortical divisions in Fig.\ref{fig:sparsity_fwds}\emph{A} alongside all subcortical divisions (Fig.\ref{fig:sparsity-inv}\emph{D}). Fig.\ref{fig:sparsity-inv}\emph{E} shows the MNE resolution matrix $\mathbf{K}^{{\rm MNE}}(\mathbf{G}_{\mathcal{B}_{r}})$. We see that the estimates for subcortical sources do not spread significantly to the cortex (Fig.\ref{fig:sparsity-inv}\emph{E}, low values for upper right and lower left). However, the estimates for subcortical sources tend to spread across the subcortical source space (Fig.\ref{fig:sparsity-inv}\emph{E}, lower right, off diagonal portions). Fig.\ref{fig:sparsity-inv}\emph{F} shows the resolution error metrics across all divisions in $\mathcal{B}_{r}$. The median SD is $2.10\,{\rm cm}$, and the median DLE is $0.805\,{\rm cm}$. This is an improvement from the previous case, but still not as accurate as needed to resolve many subcortical sources. Given a sparse cortical source space as a starting point, we anticipate that it might be possible to employ a subsequent sparse estimation step to reduce spread among subcortical sources.

\textbf{Estimation Performance - Sparse Cortical and Sparse Subcortical Sources}: Sparse estimation procedures based on $\mathit{l}_{1}$-norm minimization, projection pursuit, and/or Bayesian theory are effective at pruning out spurious features, while identifying relevant sparse features in several noisy high-dimensional problems \cite{Friedman1973,Huber1985,Needell2009,Dai2009}. We have recently shown that subspace pursuit can accurately estimate multiple sparse cortical sources underlying MEG data \cite{Babadi2014}. Thus, we assessed whether a similar subspace pursuit algorithm could reduce the spread amongst spurious subcortical sources, and enable improved resolution for subcortical source estimates.

We continue with our analysis of the resolution matrix in the composite source space $\mathcal{B}_{r}$ (Fig.\ref{fig:sparsity-inv}\emph{G}, faded background), this time with subspace pursuit. Since the subspace pursuit algorithm is non-linear, a closed-form resolution matrix in the sense of Eq.\ref{eq:resmatrix} does not exist; thus the performance of subspace pursuit must be characterized empirically instead. To this end, we simulated unit currents $\mathbf{x}_{i}$ in each $i^{{\rm th}}$ division $\mathcal{B}_{r}(i)$ within $\mathcal{B}_{r}$, one at a time, to generate corresponding noiseless field patterns $\mathbf{y}_{i}=\mathbf{G}_{\mathcal{B}_{r}(i)}\mathbf{x}_{i}$. Then, for each $\mathbf{y}_{i}$, we performed subspace pursuit and constructed an empirical resolution matrix $\mathbf{K}^{{\rm SP}}(\mathbf{G}_{\mathcal{B}_{r}})$ (see Methods). The resulting matrix, shown in Fig.\ref{fig:sparsity-inv}\emph{H}, has a near-diagonal structure for the majority of cortical and subcortical sources. Moreover, Fig.\ref{fig:sparsity-inv}\emph{I} shows SD and DLE, across all divisions in $\mathcal{B}_{r}$. The median SD and DLE are the same and equal to $0.737\,{\rm cm}$. This is a substantial improvement over previous solutions not employing sparsity constraints (Fig.\ref{fig:sparsity-inv}\emph{H} vs. Fig.\ref{fig:sparsity-inv}\emph{C}, \ref{fig:sparsity-inv}\emph{E}).

\textbf{Hierarchical Subspace-Pursuit Inverse Algorithm}:
The above results suggest that it is possible to resolve both cortical and subcortical sources by applying sparsity in both domains. In previous work, we developed a sparse estimation algorithm for cortical divisions \cite{Babadi2014}, in which sets of cortical divisions were nested in successively finer resolutions (i.e., smaller patches or divisions), and subspace pursuit was applied to derive sparse estimates in successively finer resolutions, which formed a hierarchy from coarse to fine resolution. We therefore intuited that subcortical sources could be viewed as an additional, ultimate step in this refinement process, achieved by adding a set of subcortical divisions to the final set of sparse cortical sources and applying subspace pursuit.

\begin{minipage}[b]{0.9\columnwidth}
\begin{shaded}
{\small{}\underline{\textbf{Inverse Algorithm}}}{\small \par}

{\small{}Inputs: Data $\mathbf{Y}$, distributed gain matrix $\mathbf{G}_{\mathcal{B}}$, and target sparsity level $L$. Denote the distributed cortical and subcortical source spaces as $\mathcal{B}_{C}\,{\rm and\,\mathcal{B}_{S}\,\,\subset\mathcal{B}}$ respectively. }{\small \par}
\begin{enumerate}
\item {\small{}Do subspace pursuit on distributed cortical source space $\mathcal{B}_{C}$: $[\mathcal{H}_{C},\,\hat{\mathbf{X}}_{\mathcal{H}_{C}}]={\rm SP}\left(\mathbf{Y},\,\mathbf{G}_{\mathcal{B}_{C}},\,L\right)$.}{\small \par}
\item {\small{}Construct $\mathcal{B}_{C,{\rm refined}}=\mathcal{H}_{C}\,\cup\,{\rm neighbors\,of\,\mathcal{H}_{C}}$ in a finer subdivision of cortical patches.}
\item {\small{}Repeat subspace pursuit on coarse-to-fine hierarchy of cortical source spaces $\mathcal{B}_{C,{\rm refined}}$: $[\mathcal{H}_{C_{{\rm sp}}},\,\hat{\mathbf{X}}_{\mathcal{H}_{C_{{\rm sp}}}}]={\rm SP}\left(\mathbf{Y},\,\mathbf{G}_{\mathcal{B}_{C,{\rm refined}}},\,L\right)$.}{\small \par}
\item {\small{}Construct composite space of sparse cortical sources and distributed subcortical sources: $\mathcal{B}_{\mathcal{{\rm r}}}=[\mathcal{H}_{C_{{\rm sp}}}\,\cup\,\mathcal{B}_{\mathcal{\mathcal{S}}}]$.}{\small \par}
\item {\small{}Repeat subspace pursuit on composite sparse space $\mathcal{B}_{{\rm r}}$: $[\mathcal{H}_{{\rm r}},\,\hat{\mathbf{X}}_{\mathcal{H}_{{\rm r}}}]={\rm SP}\left(\mathbf{Y},\,\mathbf{G}_{\mathcal{B}_{{\rm r}}},\,\alpha L\right)$, where $\alpha>$1.}{\small \par}
\end{enumerate}
{\small{}Outputs: Cortical and subcortical source locations $\mathcal{H}=\mathcal{H}_{{\rm r}}\subset[1,2,\ldots,K]$; and the estimated time courses of neural currents at these locations $\hat{\mathbf{X}}_{\mathcal{H}}=\mathbf{\hat{\mathbf{X}}_{\mathcal{H}_{{\rm r}}}}$.}{\small \par}
\end{shaded}
\end{minipage}

\underline{Subspace Pursuit (Steps 1, 3, and 5):} For a source space comprising a subset of brain divisions $\mathcal{F}\subset\mathcal{B}$ with gain matrices $\mathbf{G}_{\mathcal{F}}$, subspace pursuit (SP) estimates the locations and time courses of neural currents to explain data series $\mathbf{Y}_{{\scriptstyle {\scriptscriptstyle N\times T}}}$:
\begin{equation}
\mathbf{[\mathcal{H},\,\hat{\mathbf{X}}_{\mathcal{H}}]=}{\rm SP}\left(\mathbf{Y},\,\mathbf{G}_{\mathcal{F}},\,L\right)
\end{equation}
where $\mathcal{H}\subset\mathcal{F}$ denotes the set of $L$ brain divisions whose estimated neural currents $\hat{\mathbf{X}}_{\mathcal{H}}$ best explain the data $\mathbf{Y}$. Essentially, the pursuit procedure finds a sparse subset of dictionary $\mathbf{G}_{\mathcal{F}}$ that best explains data $\mathbf{Y}$, computes residuals remaining to be explained, and iterates to find new relatively uncorrelated subsets of $\mathbf{G}_{\mathcal{F}}$ that best explain these residuals, until all matching subsets of $\mathbf{G}_{\mathcal{F}}$ have been found (SI. III, \cite{Babadi2014,Dai2009,Needell2009}).

\underline{Hierarchical Construction (Steps 2 and 4):} We first apply subspace pursuit on a distributed cortical source space to identify the subset of cortical divisions that best explain the measured fields. Then, to improve accuracy of this sparse subset of cortical sources, we perform subspace pursuit across a coarse-to-fine hierarchy of cortical source spaces (SI. III, \cite{Babadi2014}). Subsequently, we construct a composite space of the sparse cortical source estimates and distributed subcortical sources; and employ subspace pursuit on this composite sparse space to identify the subset of subcortical and cortical sources that best explain the data. This process of employing sparse estimation across increasingly refined hierarchies enables systematic reduction of the distributed source space and the gain matrix: pruning out sources not important for explaining the data; implicitly decorrelating the columns of the gain matrix; and concentrating estimates into subsets of the brain whose neural currents best explain the data. Overall, given data $\mathbf{Y}$, distributed gain matrix $\mathbf{G}_{\mathcal{B}}$, and target sparsity level $L$, the hierarchical subspace pursuit algorithm identifies the sparse subset $\mathcal{H}$ that specifies locations for both cortical and subcortical sources, and estimates the time courses $\mathbf{X}_{\mathcal{H}}$ of neural currents at those locations.  

%% file: srcmod_datares.tex
\vspace{-2ex}

\section*{Data Examples}

\begin{figure*}[!ht]
\noindent \begin{centering}
\includegraphics[width=7in]{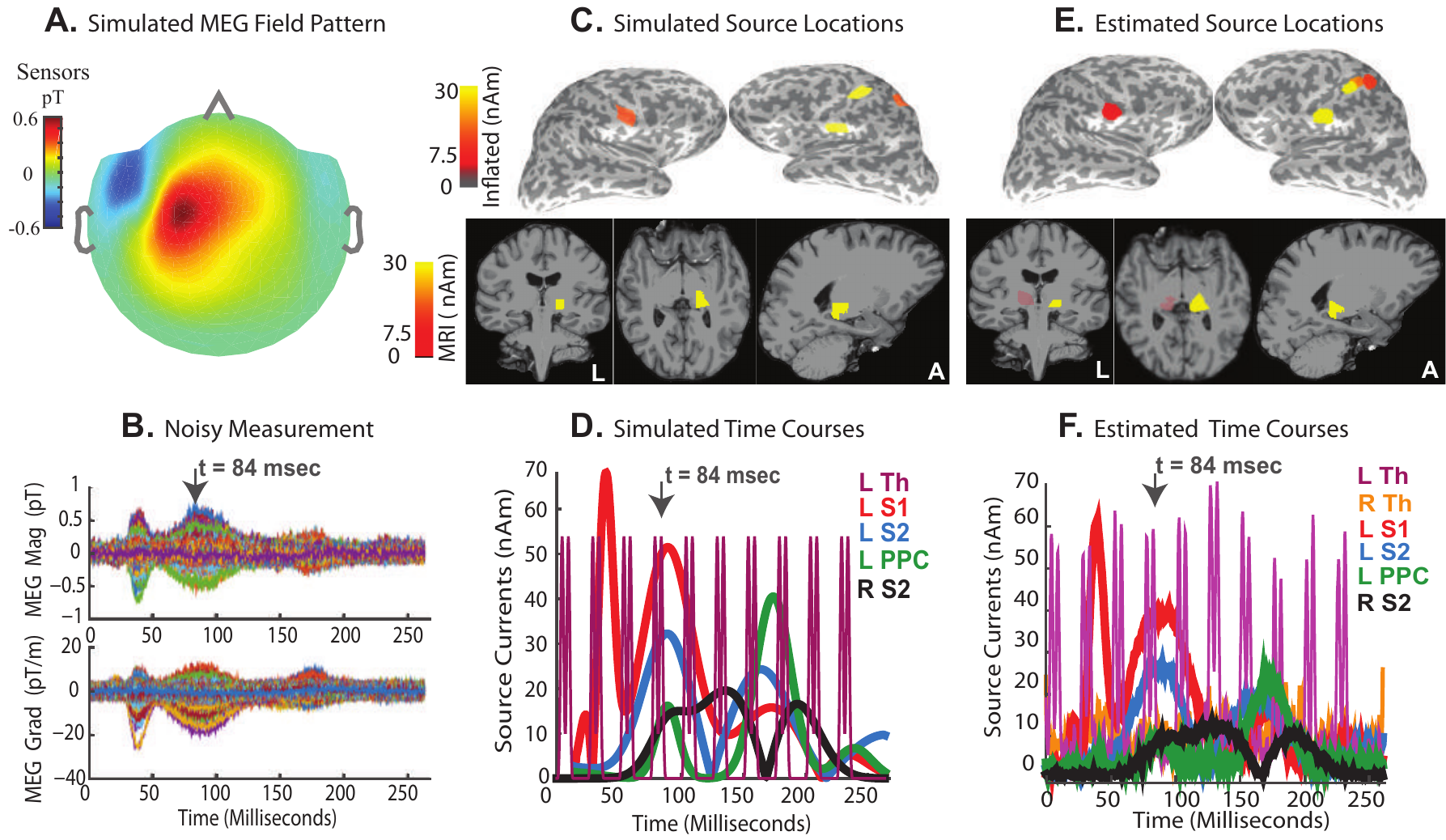}
\par\end{centering}

\caption[Sparse Hierarchical Estimates Recover Simulated Somatosensory Response]{\textbf{Sparse Hierarchical Estimates Recover Simulated Somatosensory
Response.}(\emph{A-B}) Spatial distribution and time courses (one color per channel) of simulated MEG fields in sensor space. (\emph{C-D}) Spatial distribution and time courses of simulated source currents in source space. Inflated views show sources located in somatosensory (S1, S2) and parietal (PPC) cortices. MRI views show thalamic source locations. The sagittal section passes through the left thalamus. The somatosensory thalamus (Th) is activated in a periodic on/off pattern. (\emph{D vs. B}) While the cortical source currents contribute large amplitude MEG signals, fields due to thalamic sources are not visible above the simulated noise. (\emph{E-F}) Spatial distribution and time courses of estimated source currents in the source space. All topographical snapshots are at $84\,{\rm msec}$ (top gray arrows in time course plots), the colorscales and slice locations are the same in panels \emph{C-E}, and all source currents are plotted in terms of the resultant magnitudes across dipoles within each region in the legend. \emph{(E-F vs. C-D)} Estimated source locations and time courses closely match the simulated ground truth. The thalamic source estimate follows the true phasic on/off pattern. While there is a stray source estimate in right thalamus, it is weak and relatively constant over time. SI. IV compares the performance of our algorithm to alternatives that do not employ sparsity and hierarchy.
\label{fig:sim_datares}}
\end{figure*}

We illustrate the performance of the algorithm by analyzing noisy evoked response simulations and experimental data. First, we preprocess the measurements $\mathbf{Y}$ and estimate the noise covariance matrix $\mathbf{Q}$. Next, we use the MRI data to construct the distributed source space i.e., brain divisions $\mathcal{B}$, and compute the forward solutions $\mathbf{G}_{\mathcal{B}}$. Then, we employ the hierarchical subspace-pursuit inverse algorithm to estimate the locations and time courses of sources that best explain the M/EEG data. We specify the target sparsity level $L$ for subspace pursuit empirically, based on a conservative approximation of the expected number of active divisions. We maintain $L$ across the cortical hierarchies, and increase it by a factor of $\alpha\sim1.5-2$ for the final hierarchy, i.e., the composite hybrid source space $\mathcal{B}_{r}$ comprising sparse cortical sources and distributed subcortical sources. Estimates displayed in this section are obtained at the final hierarchy level $\mathcal{B}_{r}$.

\begin{figure*}[!ht]
\noindent \begin{centering}
\includegraphics[width=7in]{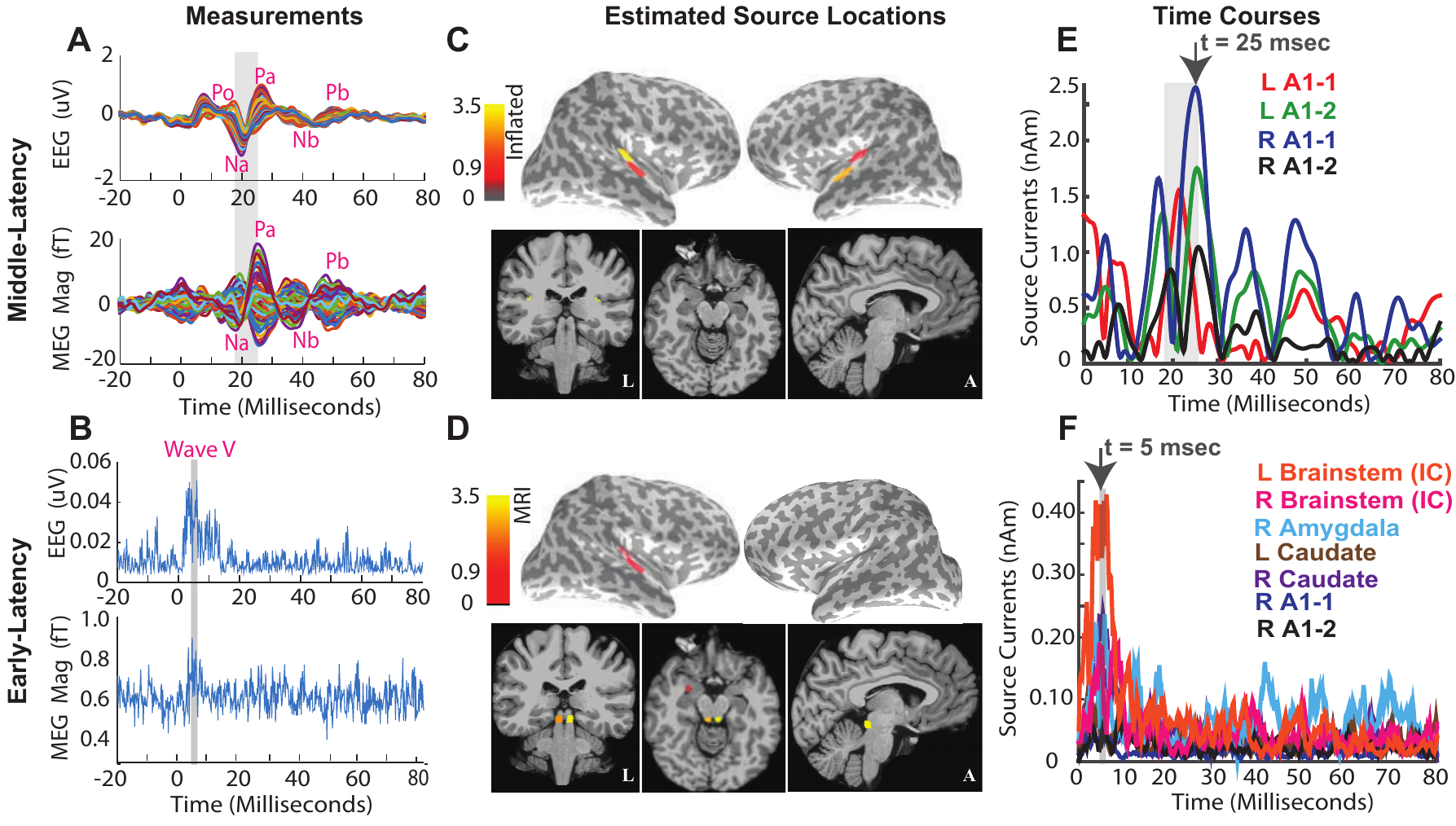}
\par\end{centering}

\caption[Cortical and Subcortical Source Estimates: Experimental Data]{\textbf{Cortical and Subcortical Source Estimates for Evoked Auditory Responses.} Stimulus locked average auditory evoked responses recorded from a healthy volunteer presented with a broadband click train stimulus. Time courses shown are averages across $11170$ epochs filtered between $500-1625\,{\rm Hz}$ for the auditory brainstem response (ABR), and $30-300\,{\rm Hz}$ for the middle latency response (MLR). (\emph{A}) MLR time courses displayed across channels, one color per channel. Red labels denote common peaks occurring at the expected post-stimulus latencies. The Na and Pa peaks (shaded gray section) are particularly prominent. \emph{(B)} ABR time courses rectified and averaged across channels. The shaded gray section marks the $5.0\,{\rm to\,6.5\,{\rm msec}}$ period post-stimulus, when peaks consistent with ABR wave V appear in the recordings. (\emph{C},\emph{E}) Sparse cortical estimates for middle-latency recordings ($30-300\,{\rm Hz)}$: snapshots at $25\,{\rm msec}$ (top gray arrow, \emph{E}). The activity is localized to the Heschl's gyrus and superior temporal gyrus, consistent with auditory cortical processing. The source time courses from these areas have peaks consistent with the Na and Pa peaks in the scalp recordings \emph{A}. \emph{(D,F)} Sparse hierarchical estimates for early-latency recordings ($500-1625\,{\rm Hz)}$, obtained using a source space comprising the sparse subsets of cortex in \emph{C} and the distributed subcortical space.  The spatial plots display the source activity at $5\,{\rm msec}$ (top gray arrow, \emph{F}). The activity is localized primarily to the inferior colliculus. A weak stray source is also seen in right amygdala. The brainstem source time courses show peaks consistent with the ABR wave V peaks in \emph{B}. The colorscales and slice locations are maintained for topographical snapshots across panels \emph{C-D}, and all source currents are resultant magnitudes across dipoles within each region in legend. The colorscales in panel \emph{C}, \emph{D} have units ${\rm nAm}$ and $0.1\,{\rm nAm}$ respectively. Overall, our sparse hierarchical algorithm recovers cortical and subcortical sources consistent with the auditory stimuli presented. SI. IV compares the performance of our algorithm to alternatives that do not employ sparsity and hierarchy.
\label{fig:abrres}}
\end{figure*}

\textbf{Somatosensory Evoked Simulations}: We simulated MEG evoked responses mimicking those elicited by electrical stimulation of the right median nerve at the wrist. Figs. \ref{fig:sim_datares}\emph{A}-\emph{B} illustrate the spatial and temporal features of simulated fields in the sensor space. The simulated fields include additive Gaussian noise (SNR $7\,{\rm dB}$ similar to resting eyes open recordings). Figs. \ref{fig:sim_datares}\emph{C-D }display the spatial and temporal patterns of simulated currents in the source space. Specifically, the evoked responses comprise early currents in the left somatosensory region of thalamus (L Som Th), followed by currents in the left primary somatosensory cortex (L S1, near the post central gyrus), and later currents in the left posterior parietal cortex (L PPC) and bilateral secondary somatosensory cortices (L/R S2). The simulated thalamic current time course has a periodic on/off pattern up to $250\:{\rm msec}$ post stimulus. This pattern was chosen to assess source estimation performance for the challenging case of phasic, temporally overlapping subcortical and cortical activity. We note that the MEG fields arising from thalamic source currents (e.g., $0-15\,{\rm msec}$ in Fig. \ref{fig:sim_datares}\emph{B}) lie below the noise floor, and are significantly smaller than those arising from cortical currents (e.g., $70-100\,{\rm msec}$ in Fig. \ref{fig:sim_datares}\emph{B}).

For source estimation, we employed a source space different from that used for the simulation, to re-create a scenario closer to what might occur in practice, in which the true generating sources and the source space parcellation might not correspond. We set the sparsity level $L$ to $8$ and $12$ for the cortical and hybrid hierarchies respectively. The procedure refines source current estimates across cortical hierarchies (SI. IV) and culminates in the final hybrid hierarchy $\mathcal{B}_{r}$. The final spatial distributions and time courses of estimated source currents (Figs. \ref{fig:sim_datares}\emph{E-F}) closely resemble those of the simulated ground truth (Figs. \ref{fig:sim_datares}\emph{C-D}) for both cortical and subcortical sources. The estimated left thalamic time course in \ref{fig:sim_datares}\emph{F} matches the simulation in shape and phase, and further is not contaminated by leakage from cortical sources.

Further, we found that our algorithm offers significant gains in performance compared to other methods that do not employ principles of sparsity and hierarchy (SI. IV). We conclude that both sparsity and hierarchy are necessary to accurately resolve locations and time courses of the thalamic source currents (SI. IV). Further, employing sparsity in a hierarchical fashion helps recover the true distribution of mean source activity across anatomic regions (SI. IV).

\textbf{Auditory Evoked Response Experiments: }We recorded simultaneous M/EEG auditory evoked responses (AEPs) elicited by binaural stimulation with a train of clicks \cite{Pockett1999,Parkkonen2009} during resting eyes-open condition. Auditory responses comprise distinct M/EEG peaks at established latencies corresponding to a progression of activity from the cochlea, through inferior colliculus in the brainstem, to the auditory cortex \cite{Pockett1999}, and thus serve as a suitable test case for validating a subcortical source estimation algorithm.

The M/EEG evoked responses are shown in Fig.\ref{fig:abrres}\emph{A}-\emph{B}. We see early ABR peaks in both EEG and MEG at $5.8\,{\rm msec}$ and $6.2\,{\rm msec}$, a low amplitude Po feature in the EEG at $\sim10\,{\rm msec}$, and prominent MLR peaks Na-Pa in MEG and EEG channels at $18-25\,{\rm msec}$ post stimulus. The ABR peaks are consistent with the brainstem Wave V known to arise from the inferior colliculus (IC), the Po feature marks the end of brainstem components, and the Na-Pa peaks correspond to cortical responses known to arise in the auditory cortex \cite{Pockett1999}.

We performed hierarchical subspace pursuit, and set sparsity levels $L$ to $4$ and $8$ for the cortical and hybrid hierarchies respectively. Figs.\ref{fig:abrres}\emph{C, E} show localization of the $18-25\,{\rm msec}$ Na-Pa MLR peaks to bilateral auditory cortices, and the associated time courses. The auditory areas comprise the reduced cortical source space, which along with the distributed subcortical sources, form the hybrid source space $\mathcal{B}_{r}$ for estimation of deep sources underlying the ABR data. Fig.\ref{fig:abrres}\emph{D,F} illustrate the localization of the Wave V ABR peaks to the IC. Although the recorded Wave V peaks do not have very high SNR, the source time course at IC peaks at $\sim5-6\,{\rm msec}$ and drops off after $10\,{\rm msec}$, as expected. We compared performance to algorithms that do not employ sparsity and hierarchy (SI. IV), and found that hierarchical subspace pursuit is necessary to estimate specific subcortical sources even for filtered recordings containing temporally separated early latency responses. 

%% file: srcmod_discussion.tex
\vspace{-2ex}

\section*{Discussion}
The extent to which subcortical activity can be estimated from M/EEG measurements is controversial. Our key finding is that the M/EEG fields from subcortical sources can be distinguished from those generated by the cortex when the underlying cortical activity is sparse, a condition that is relevant in many neuroscience investigations. In this scenario, the problem of distinguishing subcortical from cortical sources has a similar level of ambiguity as that of resolving different cortical sources. We devised a source estimation algorithm that takes advantage of this insight by estimating both sparse cortical and subcortical sources in a hierarchical fashion.

It is known that deep and superficial sources exhibit different M/EEG field patterns \cite{Williamson1981}, but the degree to which this information could be used to resolve multiple distributed subcortical and cortical sources has remained unclear. Analyses of cortical and subcortical field patterns assuming that entire structures can be simultaneously active have provided evidence for substantial correlation \cite{Attal2007}, consistent with our Fig.\ref{fig:challenge}. However, we reasoned that it would be unlikely to observe synchronous activity, the major determinant of MEG/EEG \cite{BookChHari2002}, simultaneously within the entirety of cortex or any given subcortical structure. Thus, we analyzed sparse subdivisions of cortical and subcortical structures, and found clear distinctions in the ensuing field patterns (Fig.\ref{fig:sparsity_fwds}). Although previous work and the data presented here show that resolution of cortical and subcortical sources is fundamentally ambiguous, we found that if the distribution of cortical and subcortical sources is sparse, the problem becomes tractable.

These observations motivated us to create a hierarchical subspace pursuit inverse algorithm to find the set of sparse cortical and subcortical sources, which best explain the M/EEG data. Our analyses of various source estimators (Fig.\ref{fig:sparsity-inv}) showed our algorithm has a performance superior to alternatives for the subcortical structures and similar to existing approaches for the cortical structures \cite{Nunez1995,Mosher1993,Sekihara2005,NunezPaul2006,Babadi2014}.

If the locations of activity in cortical and subcortical structures are known, and each active area can be modeled with an equivalent current dipole, linear least squares can be employed to estimate source current time courses \cite{Tesche1996,Tesche1997}. However, fitting the locations of multiple dipolar sources usually requires tailored and often interactively guided fitting strategies. Our approach, on the other hand, automatically finds the constellation of sources in a variety of conditions, including those where the source activities may overlap in time. Further, instead of isolated current dipoles we employ concise distributed dipole representations within relevant subcortical anatomical subdivisions. Other methods, such as Magnetic Field Tomography (MFT) \cite{Ioannides1995} and the linearly constrained minimum-variance (LCMV) beamformer \cite{beamformer2008}, that have been applied to locate deep sources implicitly employ some of the principles we formalize here. Our analyses on simulated and experimental data show that algorithms employing both sparsity and hierarchy can resolve simultaneously active cortical and subcortical sources under low SNR conditions (Figs.\ref{fig:sim_datares}-\ref{fig:abrres}).

Several recent publications have proposed sparsity-based algorithms for M/EEG cortical source estimation \cite{Gorodnitsky1995,Mosher1998,Uutela1999,Durka2005,Friston2008,Vega-Hernandez2008,Ou2009,Wipf2010, Gramfort2011,Gramfort2012,Gramfort2013,Babadi2014}. Based on our insights, we suggest that these other methods could be adapted and extended to provide a variety of practical options for estimating subcortical sources based on M/EEG data. Further, techniques employing distributed sparse representations and dynamical sparsity constraints could be used to estimate subcortical sources in conditions involving more extended areas of cortex.

In conclusion, we achieved fundamental new insights based on the biophysics of the M/EEG inverse problem and the anatomical locations of the relevant brain structures. Based on this, we developed practical tools to estimate simultaneous cortical and subcortical activity from M/EEG measurements. We demonstrated the efficacy of our source estimation algorithm by analyzing both realistic simulations and experimental data. Our analysis and method provide new opportunities for characterizing electrophysiological dynamics in subcortical structures of the human brain. 

%% file: srcmod_methods.tex
\vspace{-2ex}

\section*{Methods}

\small{
\textbf{{\normalsize{}MRI and M/EEG Acquisition}}
We acquired MRI and M/EEG data on 5 healthy subjects aged $25-45\,{\rm years}$. The subjects provided written informed consent. All studies were approved by the Human Research Committee at Massachusetts General Hospital. We obtained T1-weighted structural MRI (Siemens 3T TimTrio${\rm ^{TM}}$, multi-echo MPRAGE, TR = $2,510\,{\rm ms}$; 4 echoes with TEs = $1.64$, $3.5$, $5.36$ and $7.22\,{\rm ms}$; $176$ sagittal slices, $1-{\rm mm}$ isotropic voxels, $256\times256$ matrix; flip angle = $7^{\circ}$). We acquired eyes-open MEG and EEG data using a $306$-channel Elekta-Neuromag Vectorview array (Helsinki, Finland) and a $70$-electrode EEG cap. We registered the subject's head position relative to the MEG sensors using head-position indicator (HPI) coils. We digitized the locations of the HPI coils, EEG electrodes, and the scalp using a FastTrak 3D digitizer (Polhemus, Inc., VT, USA), and aligned these locations with the MRI using the MNE software package \cite{Hamalainen1989,AlexanderGramfortM.Luessi}.

\textbf{{\normalsize{}Source Space Construction}}
We used FreeSurfer to reconstruct neocortical and hippocampal surfaces, and segment subcortical volumes from the MRI \cite{Dale1999,Fischl1999,Fischl2002}. We placed dipoles with orientations normal to the triangulated surface mesh for neocortex and hippocampus at the gray-white matter interface, with $\sim1\,{\rm mm}$ spacing. We placed triplets of orthogonal dipoles in subcortical volumes covering the thalamus, putamen, caudate, brainstem (midbrain), and amygdala, at $\sim1\,{\rm mm}$ voxel spacing. To reduce dimensionality of the source space, we grouped neighboring cortical dipoles into ``patches'' \cite{Hamalainen1989,AlexanderGramfortM.Luessi,Limpiti2006}. We grouped neighboring subcortical dipoles into ``subdivisions'' sized to produce signals with similar amplitudes as the cortical patches (SI. I). For cortical patches with an average area of $175\,{\rm mm^{2}}$, this sizing procedure resulted in $209$ subcortical subdivisions with volumes ranging $175-1800\,{\rm mm^{3}}$ (see Fig.\ref{fig:srcspace}\emph{B}). Regions with higher current density have finer divisions (higher resolution) than those with low current density (e.g., $\sim200\,{\rm mm^{3}}$ divisions in striatum vs. $\sim1795\,{\rm mm^{3}}$ divisions in thalamus).

\textbf{{\normalsize{}Forward Solutions}}
We derived a three-compartment boundary-element model from the MRI data and numerically computed forward solutions using the MNE software package \cite{Hamalainen1989,AlexanderGramfortM.Luessi}. To account for the different sensor types, units, and noise levels in the M/EEG measurements, we whitened the gain matrices using an estimate of the observation noise covariance matrix \cite{AlexanderGramfortM.Luessi}. For MEG simulation studies, we set the noise covariance matrix to be similar to typical resting eyes open recordings: $\mathbf{Q}={\rm diag([g^{2},\,g^{2},\,m^{2},\,\ldots,g^{2},\,g^{2},\,m^{2}])}$ where ${\rm g}=2.5\,{\rm fT/cm}$ and ${\rm m}=10\,{\rm fT}$. For M/EEG experimental studies, we estimated noise covariance matrices from the resting eyes-open data. To account for differences in current strength across brain divisions, we scaled gain matrices for each division by the regional current strengths (\cite{Attal2012}, SI.I). We constructed the reduced dimensionality M/EEG gain matrices $\mathbf{G}_{k}$ for each division $k\,\epsilon\,\{\mathcal{B}:\,1,\,2,\,\ldots,\,K\}$  using a singular value decomposition, retaining components to capture $>95\%$ of the total spectral energy \cite{Limpiti2006,Babadi2014}.

\textbf{{\normalsize{}Analysis of Forward Solutions: Field Patterns and Principal Angles}}
We simulated the field pattern $\mathbf{y}_{k}$ for a division $k$ by activating the most significant eigenmode of $\mathbf{G}_{k}$ with a unit current. We assessed the degree to which the field pattern $\mathbf{y}_{k}$ could be explained by some distribution of currents in a region $\mathcal{R}\subset\mathcal{B}$ $(k\,\notin\,R$) by fitting $\mathbf{G}_{\mathcal{R}}\mathbf{x}_{\mathcal{R}}$ to $\mathbf{y}_{k}$ using the regularized $l_{2}$ minimum-norm criterion. We computed $\hat{\mathbf{x}}_{\mathcal{R}}=\mathbf{G}_{\mathbb{\mathcal{R}}}^{'}\left(\mathbf{G}_{\mathbb{\mathcal{R}}}\mathbf{G}_{\mathcal{R}}^{'}+\lambda^{2}\mathbf{I}\right)^{-1}\mathbf{y}_{k}$, setting the regularization parameter $\lambda^{2}$ to $1/9$. We then quantified the goodness of fit between the best fitting field patterns, $\mathbf{\hat{\mathbf{y}}}_{\mathcal{R}}=\mathbf{G}_{\mathcal{R}}\hat{\mathbf{x}}{}_{\mathcal{R}}$ and the original pattern $\mathbf{y}_{k}$, as: $1-||\mathbf{y}_{k}-\mathbf{\hat{\mathbf{y}}}_{\mathcal{R}}||_{2}/||\mathbf{y}_{k}||_{2}$. For visualization of the fields, we dewhitened $\hat{\mathbf{y}}_{\mathcal{R}}$ and mapped it to a virtual grid of $304$ magnetometers distributed evenly across the Elekta helmet \cite{HamalainenMattiandIlmoniemi1994}.

We used principal angles \cite{Bjorck1973,Knyazev2008} to quantify the correlation between putative field patterns arising from currents in two non-overlapping regions: $\mathcal{R}_{1}\subset\mathcal{B}$ and $\mathcal{R}_{2}\subset\mathcal{B}$. Any putative field pattern arising from a current distribution in $\mathcal{R}_{1}$ is defined by some subset $\mathbf{U}_{\mathcal{R}_{1}}$ of the eigenmodes of the gain matrix $\mathbf{G}_{\mathcal{R}_{1}}$. If $\mathbf{G}_{\mathcal{R}_{1}}$ comprises $n1$ eigenmodes, there are $2^{n1}-1$ subsets of eigenmodes $\{\mathbf{U}_{\mathcal{R}_{1}}\}$ describing possible current patterns within $\mathcal{R}_{1}$. Similarly, for region $\mathcal{R}_{2}$, the gain matrix $\mathbf{G}_{\mathcal{R}_{2}}$ comprises $n2$ eigenmodes, and there are $2^{n2}-1$ subsets of eigenmodes $\{\mathbf{U}_{\mathcal{R}_{2}}\}$ describing possible current patterns within $\mathcal{R}_{2}$. The degree to which some field pattern arising from a current in $\mathcal{R}_{1}$ can be explained by a field arising from some current in $\mathcal{R}_{2}$ is specified by the set of $(2^{n2}-1)$ principal angles $\{\Theta_{1,2}\}$ between $\mathbf{U}_{\mathcal{R}_{1}}$and each of the $(2^{n2}-1)$ subsets in $\{\mathbf{U}_{\mathcal{R}_{2}}\}$. We computed angles $\{\Theta_{1,2}\}$ across all possible current distributions in $\mathcal{R}_{1}$, i.e., for each of the $(2^{n1}-1)$ subsets in $\{\mathbf{U}_{\mathcal{R}_{1}}\}$. The distribution of these $(2^{n1}-1)\,(2^{n2}-1)$ sets of principal angles characterizes the correlation between any field patterns that could arise from currents in $\mathcal{R}_{1}$ and $\mathcal{R}_{2}$. To illustrate how the principal angles correspond to different fields with varying levels of similarity, we selected a representative combination of eigenmodes $\mathbf{U}_{\mathcal{R}_{1}}$ and $\mathbf{U}_{\mathcal{R}_{2}}$ having angles $35^{\circ}$, $45^{\circ}$, $60^{\circ}$and $86{}^{\circ}$, generated the field $\mathbf{y}_{1}=\mathbf{U}_{\mathcal{R}_{1}}$, projected (fit with $l_{2}$ minimum-norm as above) $\mathbf{y}_{2}=\mathbf{U}_{\mathcal{R}_{2}}\mathbf{x}_{\mathcal{R}_{2}}$ to $\mathbf{y}_{1}$, and plotted the $\mathbf{y}_{2}$ most similar to $\mathbf{y}_{1}$.

\textbf{{\normalsize{}Analysis of Resolution Matrices for Inverse Solutions}}
We used the resolution matrix to assess performance of the minimum $l_{2}$-norm (MNE) and subspace pursuit (SP) inverse solutions. The MNE solution for the distributed source space $\mathcal{B}$, with pre-whitened gain matrix $\mathbf{G}_{\mathcal{B}}$, is: $\mathbf{W}^{{\rm MNE}}\left(\mathbf{G}_{\mathcal{B}}\right)\,=\mathbf{\,RG_{\mathcal{B}}^{'}\mathbf{\left(G_{\mathcal{B}}R\mathbf{G_{\mathcal{B}}^{'}+}\lambda^{2}I\right)^{-1}}}$. We specified the prior source covariance $\mathbf{R}$ so that ${\rm Tr(\mathbf{G}\mathbf{_{\mathcal{B}}R}\mathbf{G}_{\mathcal{B}}^{'})=Tr(\mathbf{I})}=N$, set $\lambda^{2}=1/9$ \cite{HamalainenMattiandIlmoniemi1994}, and computed the MNE resolution matrix $\mathbf{K}^{{\rm MNE}}(\mathbf{G}_{\mathcal{B}})=\mathbf{W}^{{\rm MNE}}(\mathbf{G_{\mathcal{B}}})\mathbf{G}_{\mathcal{B}}$ \cite{Liu1998}. We used a similar process to compute the MNE resolution matrix $\mathbf{K}^{{\rm MNE}}(\mathbf{G}_{\mathcal{B}_{r}})$. The subspace pursuit (SP) estimates for the source space $\mathcal{B}_{r}$, with pre-whitened gain matrix $\mathbf{G}_{\mathcal{B}_{r}}$, were obtained using the procedure in SI. III. We characterized the SP resolution matrix $\mathbf{K}^{{\rm SP}}$ empirically. We simulated unit currents in the most significant eigenmode of each brain division $i$, and used the gain matrix $\mathbf{G}_{\mathcal{B}_{r}(i)}$ to generate the noiseless (pre-whitened) MEG fields $\mathbf{y}_{i}$. We then used subspace pursuit to estimate the source location $q_{i}={\rm SP(\mathbf{y}_{i},}\,\mathbf{G}_{\mathcal{B}_{r}},\,L=1)$, and computed the source current $\beta_{\{q_{i},\,i\}}$ using a least squares fit of $\mathbf{G}_{\mathcal{B}_{r}(q_{i})}$ to $\mathbf{y}_{i}$. The estimated $q_{i}$ and $\beta_{\{q_{i},\,i\}}$ together specify the locations and magnitudes of the non-zero elements of the empirical resolution matrix $\mathbf{K}^{{\rm SP}}$. As $L=1$, only the $q_{i}^{{\rm th}}$ element of column $i$ of $\mathbf{K}^{{\rm SP}}$ is non-zero. We used the resolution matrices $\mathbf{K}^{{\rm {\rm MNE}}}$ and $\mathbf{K}^{{\rm SP}}$ to compute the spatial dispersion ${\rm SD}_{i} = \sqrt{\sum_{k=1}^{N}(d_{ki}\cdot||\mathbf{K}_{ki}||)^{2}/\sum_{k=1}^{N}||\mathbf{K}_{ki}||{}^{2}}$ and the dipole localization error ${\rm DLE}_{i}=d_{ji}$, where $j=\arg\max_{k}\{||\mathbf{K}_{ki}||\}$ and $d_{ji}$ is the distance between centroids of divisions $j$ and $i$ \cite{Molins2008}.

\textbf{{\normalsize{}Source Estimation Algorithm}}
At each hierarchy level, we performed source estimation using a modified version of the subspace pursuit (SP) algorithm described in \cite{Babadi2014}. This algorithm adapted the original SP algorithm \cite{Dai2009,Needell2009} by employing: (a) the MNE proxy, in lieu of the standard projection, for selecting the sparse subsets of a given dictionary that explain the data; and (b) a mutual coherence threshold of $\sim0.5$ reflecting the mean-max correlation amongst gain matrices from random pairs of cortical patches, in lieu of the restricted isometry property, to specify the maximum correlation allowed between the chosen sparse subsets.  In this work, we relaxed the mutual coherence threshold to be the mean-max correlation amongst gain matrices from neighborhoods of brain divisions under consideration. The relaxed mutual coherence allows the threshold to adapt to changing levels  of gain matrix correlation across hierarchy levels, giving correlation thresholds in the $\sim0.75-0.90$ range. Additional details are provided in SI.III.

\textbf{{\normalsize{}Visualization of Source Estimates}}
For each brain division $h\,\epsilon\,\mathcal{H}$, we converted the estimated eigenmode current time courses $\hat{\mathbf{X}}_{h}$ to the elementary dipole current time courses, accounting for the regional current strength normalization used to construct the gain matrices. We then computed a vector sum of these dipole currents to derive a resultant time course $\hat{\mathbf{c}}_{h}$, denoting the estimated current magnitude in a representative division $h$ as a function of time. We illustrated spatial distributions plotting $\hat{\mathbf{c}}_{\mathcal{H}}$ at a time point of interest on inflated surfaces and MRI slices. We plotted the time courses $\hat{\mathbf{c}}_{\mathcal{H}}$ to show the temporal evolution. We summarized the spatial distribution of estimates in bar graphs plotting the root mean square magnitudes (across time) of estimated currents as a function of anatomical region $\mathcal{R}$: ${\rm mean_{h\,\epsilon\,\mathcal{H}_{{\rm {\rm \mathcal{R}}}}}}\left(\sqrt{||\hat{\mathbf{c}}_{h}||_{{\scriptstyle 2}}}\right)$.

\textbf{{\normalsize{}Somatosensory Evoked Response Simulations}}
We simulated activity in five regions of interest: a $1\,{\rm cm^{3}}$ volume in the left somatosensory thalamus including the ventral posterior area, and 4 $\sim600-800\,{\rm mm^{2}}$ surface patches in primary and secondary somatosensory cortices and the posterior parietal area. The simulated cortical current time courses were Gabor atoms of the form $Ae^{-(t-t_{o})/2\sigma^{2}}$, where $A$, $t_{o}$ and $\sigma$ denote the amplitude, delay and width of the evoked component, and were set based on previous studies \cite{Gramfort2011,Babadi2014,NeidermeyerErnstDaSilva}. The simulated thalamic current time course consisted of $10$ repetitions of \textbf{ }$\cos^{2}(2*\pi*f*t+\phi)$ with $f=100\,{\rm Hz}$, $\phi=\pi/3$, duration $15\,{\rm msec}$ and repetition period $25\,{\rm msec}$. We simulated currents over a time range of $0-250\,{\rm msec}$, with $3\,{\rm {\rm kHz}}$ sampling rate.  We calculated the MEG signals using the MRI-based forward model, and added white Gaussian observation noise with mean zero and variance specified to achieve a SNR of $\sim7\,{\rm dB}$.

\textbf{{\normalsize{}Auditory Evoked Response Recordings}}
We measured auditory evoked potentials and fields in two healthy subjects. We used Presentation$^{{\rm TM}}$ software (v17.1, Neurobehavioral Systems, Inc., Albany, CA, USA) to present trains of binaural broadband clicks ($0.1\,{\rm msec}$ duration, $65-80\,{\rm dB/nHL}$ intensity, $110\,{\rm msec}$ inter-stimulus interval,$9.09\,{\rm Hz}$ click rate) as the subjects rested with eyes open \cite{Makela1994,Parkkonen2009}. We recorded M/EEG data simultaneously at a sampling rate of $5\,{\rm kHz}$, bandpass filtered between $0.03$ and $1660\,{\rm Hz}$. We recorded $2\,{\rm {\rm min}}$ each of empty room MEG and pre-stimulus baseline, followed by $5$ evoked potential runs ($5.5\,{\rm min}$ each) to obtain  $>10000$ epochs for averaging. To account for ocular and electrocardiographic artifacts, we also recorded electrooculograms (EOG) and electrocardiograms (ECG) during the study.

We preprocessed the raw M/EEG data to remove power line noise (comb notch filter, MATLAB$^{{\rm TM}}$); and excluded artifactual channels (marked by inspection) and eye-blink epochs (peak-peak EOG$>150\,{\rm uV}$ in $1-40\,{\rm Hz}$ band). We bandpass filtered the preprocessed data to $500-1625\,{\rm Hz}$, and $30-300\,{\rm Hz}$ to obtain the early auditory brainstem response (ABR) and the middle latency response (MLR) components of the auditory evoked potential, respectively \cite{Makela1994,Parkkonen2009,Pockett1999}. We used signal space projection (SSP) to attenuate environmental noise \cite{AlexanderGramfortM.Luessi}. We computed stimulus-locked averages for the MLR and ABR, correcting for the sound tube delay of $9.775\,{\rm msec}$. We estimated MLR and ABR observation noise covariances using the filtered baseline eyes open recordings (\cite{Parkkonen2009}, SI. IV), and whitened the filtered MEG and EEG measurements using their respective noise covariances. We estimated sources using both MEG and EEG for each of the MLR and ABR bands.
}